\documentclass[12pt,number]{elsarticle}
\usepackage{graphicx}
\usepackage{amsmath}
\usepackage{amsfonts}
\usepackage{amssymb}
\usepackage{longtable}
\usepackage{array}

\begin{document}

\journal{Phys. Lett. A}

\begin{frontmatter}
\title{The nuclear structure and related properties of some low-lying isomers of
free-space O$_n$ clusters ($n=6$, $8$, $12$)}
\author[CTChem]{G. Forte}
\author[CTPhys,SSC,CNISM,INFN]{G. G. N. Angilella\corref{corr}}
\ead{giuseppe.angilella@ct.infn.it}
%\author[CTChem]{V. Pittal\`a}
\author[AntwerpPhys,Oxford]{N. H. March}
\author[CTPhys,CNISM]{R. Pucci}

\address[CTChem]{Dipartimento di Scienze del Farmaco, Universit\`a di Catania,\\
Viale A. Doria, 6, I-95126 Catania, Italy}
\address[CTPhys]{Dipartimento di Fisica e Astronomia, Universit\`a di Catania,\\
Via S. Sofia, 64, I-95123 Catania, Italy}
\address[SSC]{Scuola Superiore di Catania, Universit\`a di Catania,\\ Via
Valdisavoia, 9, I-95123 Catania, Italy}
\address[CNISM]{CNISM, UdR Catania, Via S. Sofia, 64, I-95123 Catania, Italy}
\address[INFN]{INFN, Sez. Catania, Via S. Sofia, 64, I-95123 Catania, Italy}
\address[AntwerpPhys]{Department of Physics, University of Antwerp,\\
Groenenborgerlaan, 171, B-2020 Antwerp, Belgium}
\address[Oxford]{Oxford University, Oxford, UK}
\cortext[corr]{Corresponding author.}

\begin{abstract}

After some introductory comments relating to antiferromagnetism of crystalline
O$_2$, and brief remarks on the geometry of ozone, Hartree-Fock (HF) theory plus
second-order M\o{}ller-Plesset (MP2) corrections are used to predict the nuclear
structure of low-lying isomers of free-space O$_n$ clusters, for $n=6$, $8$,
and $12$. The equilibrium nuclear-nuclear potential energy is also discussed in
relation to the number $n$ of oxygen atoms in the cluster.

\end{abstract}

\end{frontmatter}

%\section{Introduction}
%\label{sec:introduction}

As is well known, the O$_2$ molecule has parallel spin electrons in its ground
state. This results in important magnetic correlations between
molecular or cluster units in oxygen assemblies, especially at high density,
giving rise to a variety of `spin-controlled' solid phases of molecular oxygen
under pressure \cite{Freiman:04}. A remarkable result of the triplet character
of oxygen molecules is the existence of insulating crystalline O$_2$ in an
antiferromagnetic (AF) state \cite{Goncharenko:04}. This, to our knowledge, is
the only AF insulator known in nature with a single atomic species present,
though metallic AFs exist in the forms of Cr and Mn. Spin ordering has
relevant consequences on  electronic transport properties in solid oxygen. At
high pressures, solid oxygen changes from an insulating to a metallic state
\cite{Desgreniers:90}, and at sufficiently low temperatures it even undergoes a
superconducting transition \cite{Shimizu:98}. In the so-called red- or
$\varepsilon$-phase \cite{Nicol:79,Akahama:95}, high-pressure solid oxygen is
believed to consist of O$_n$ clusters in a crystal lattice, with $n=4$
\cite{Gorelli:99} or $n=8$ \cite{Fujihisa:06}. As to free-space ozone, the
nuclear structure takes the form of an isosceles triangle, though we do not know
of any crystalline form, even under high pressure, for which O$_3$ is the basic
building block. Turning more briefly to O$_4$, at least two theoretical studies
of its low-lying free-space isomers exist \cite{Seidl:92,Caffarel:07}.

Here, we shall use quantum-chemical methods, first of all, to study the nuclear
structures of some low-lying isomers of the free-space clusters O$_n$, for
$n=6$, $8$, and $12$. The problem of determining the free-space
structure and other dynamical properties of oxygen clusters has been seldom
addressed in the literature, one exception being Ref.~\cite{Gimarc:94}, where
however the geometry was restricted to the ring shape. Here, Hartree-Fock plus
second-order M\o{}ller-Plesset (HF+MP2) theory \cite{Moeller:34} was applied
first of all for the optimization of the clusters with subsequent additions of
more refined methods. All calculations were  performed by means of Gaussian~09
software \cite{Frisch:09a}, where the 6-311 basis set
\cite{Mclean:80,Krishnan:80} was used, supplemented by polarization and diffuse
functions (6-311++G**).

\begin{figure}[t]
\centering
\includegraphics[width=0.45\columnwidth]{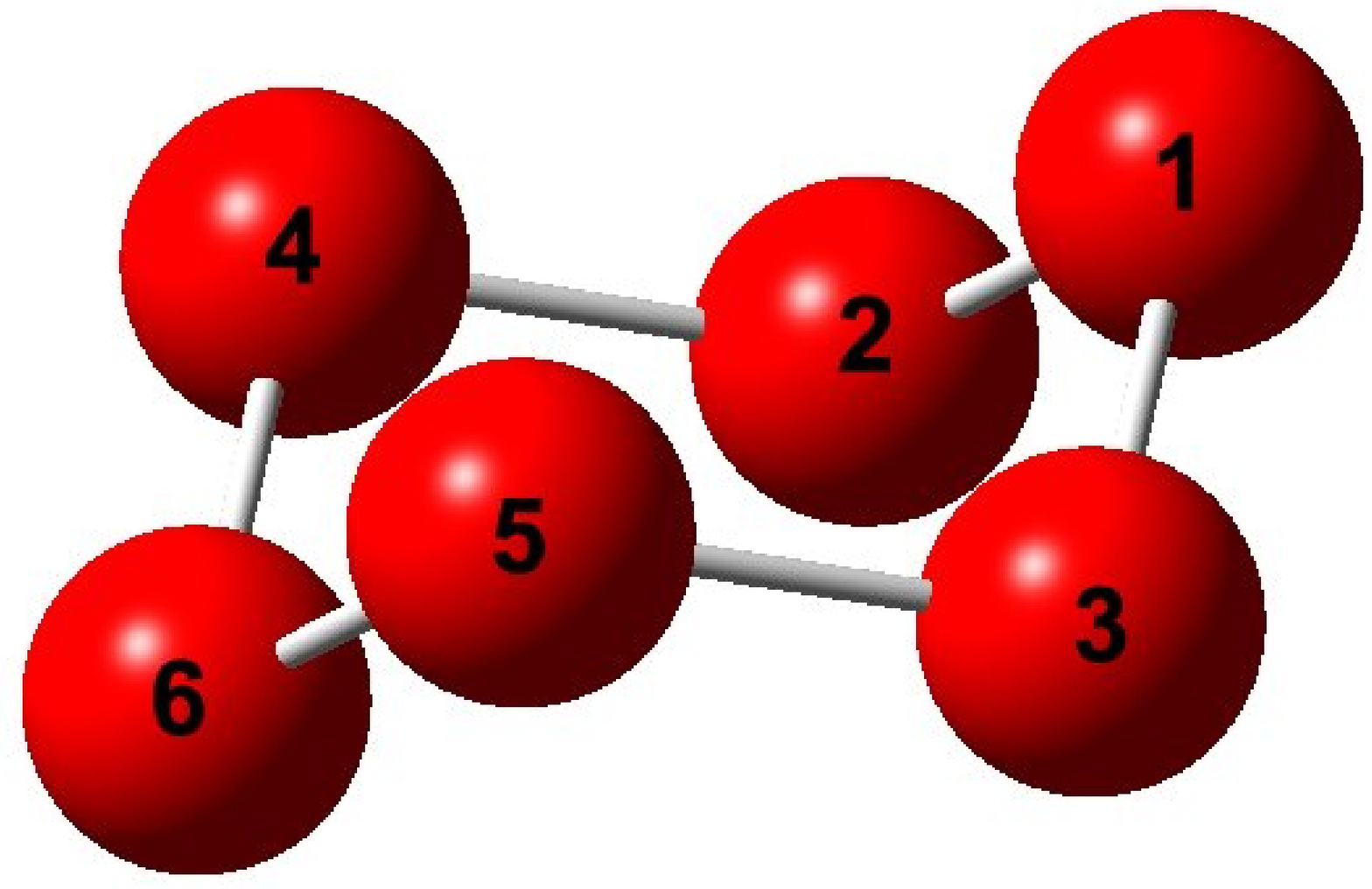}
\caption{Low-lying isomer of O$_6$. See Table~\ref{tab:structure} for structural
details, and Table~\ref{tab:energies} for ground-state and dissociation energy.}
\label{fig:1}
\end{figure}

Fig.~\ref{fig:1} shows a low-lying isomer of O$_6$, which appears as a linkage
between two somewhat perturbed, of course, ozone molecules. Table~\ref{tab:structure}
collects the various bond lengths and relevant angles associated with this
nuclear structure. The ground-state energy is given in atomic units (a.u.),
together with the dissociation energy into six O atoms.

Next, we show in the left panel of Fig.~\ref{fig:2} a predicted nuclear
structure of O$_8$, in the singlet ($s$) state, by optimization using HF+MP2
theory. Calculations within density functional theory (DFT) using 
the functional of Perdew, Burke, and Ernzerhof (PBE) \cite{Perdew:96}, which has
been shown by Amovilli \emph{et al.} \cite{Amovilli:09} to have possibly
variational validity, lead to a related structure, shown in the right panel of
Fig.~\ref{fig:2} (see Tab.~\ref{tab:structure} for structural details). However
(see further comments below), the possibility that this structure represents a
transition state cannot be ruled out, as that may require really careful
treatment of electron correlation.

\begin{figure}[t]
\centering
\includegraphics[width=0.5\columnwidth]{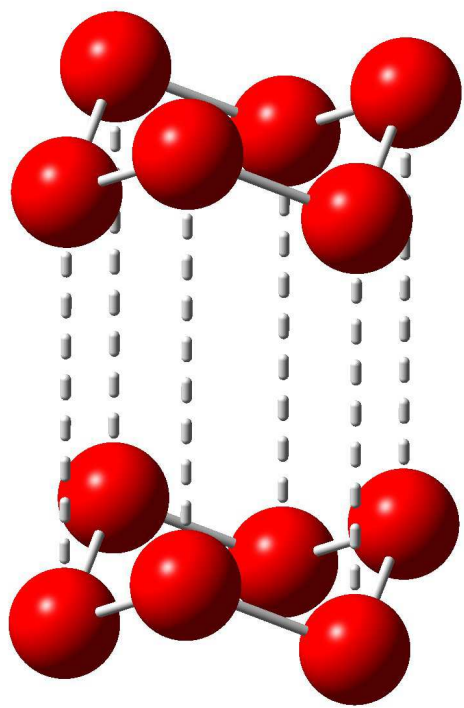}
\caption{Low-lying isomer of O$_{12}$. This is essentially composed by two
O$_6$, weakly bound sub-clusters, separated by an average distance of
$3.37$~\AA. Within each sub-cluster, one finds alternating angles of
$103.90^\circ$ and $104.00^\circ$ between two consecutive $O--O$ bonds.
One also finds an overall singlet spin configuration.
See Table~\ref{tab:energies} for ground-state and dissociation energy.}
\label{fig:3}
\end{figure}

Fig.~\ref{fig:3} depicts what we find to be a low-lying isomer of O$_{12}$. 

\begin{figure}[t]
\centering
\includegraphics[height=0.8\columnwidth,angle=-90]{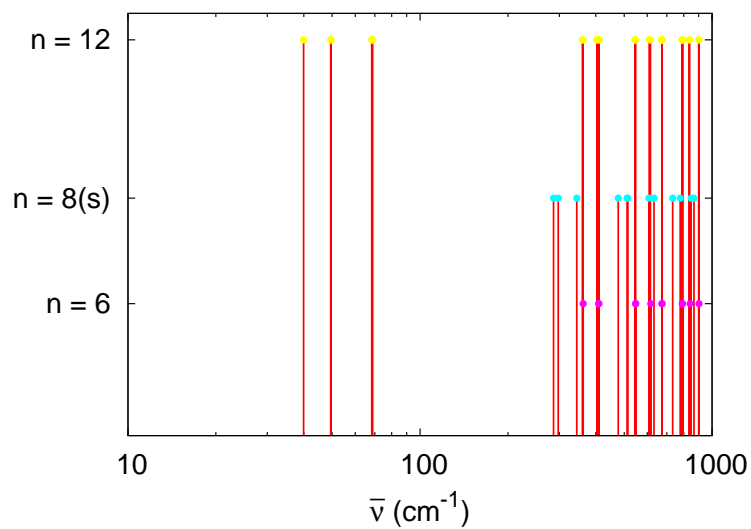}
\caption{Calculated frequencies of the normal modes of the three clusters under
study.}
\label{fig:frequencies}
\end{figure}

Fig.~\ref{fig:frequencies} shows the calculated frequencies of the normal modes
for the different oxygen clusters under consideration. While all frequencies are
real in the case $n=6$ (Fig.~\ref{fig:1}), some frequencies turn out to be
imaginary in the case $n=8$. 
This has been confirmed by further calculations at the CCSD level of
approximation (employing the 6-311++** basis set). Specifically, while the O$_8$
cluster in the triplet configuration does not form at all, one finds a
O$_8$ cluster in the singlet configuration, with a geometry much similar to the
one shown in the left panel of Fig.~\ref{fig:2} (see Tab.~\ref{tab:structure} for
structural details). However, two frequencies turn out to be imaginary, thus
showing that this structure might correspond to a second order saddle point in
the configurational energy (null gradient, but two imaginary frequencies).

Then, as a further related property, we have also extracted from the present
bond lengths, the bare electrostatic nuclear-nuclear potential energy, $U_{nn}$,
for all the low-lying isomers considered here. This is stimulated by the early
work of Mucci and March \cite{Mucci:79}, who studied the equilibrium values of
$U_{nn}$ for a wide variety of tetrahedral and octahedral molecules. Their
surprising conclusion was that $U_{nn}$ correlated closely with the total number
of electrons in these molecules. Prompted by this finding, and its relevance to
density functional theory (DFT) \cite{Parr:89}, we have plotted in
Fig.~\ref{fig:4} $U_{nn}$ \emph{vs} the number $n$ of oxygen atoms
for the four clusters considered above.

In summary, our main findings are for the nuclear arrays proposed in
Figures~\ref{fig:1}--\ref{fig:3} for isomers of free-space O$_n$
clusters ($n=6,8,12$). An additional finding is that the equilibrium
nuclear-nuclear potential energy of these predicted structures accord with the
conclusions reached by Mucci and March \cite{Mucci:79} on tetrahedral and
octahedral molecules.

\section*{Acknowledgements}

\begin{figure}[t]
\centering
\includegraphics[width=0.45\columnwidth]{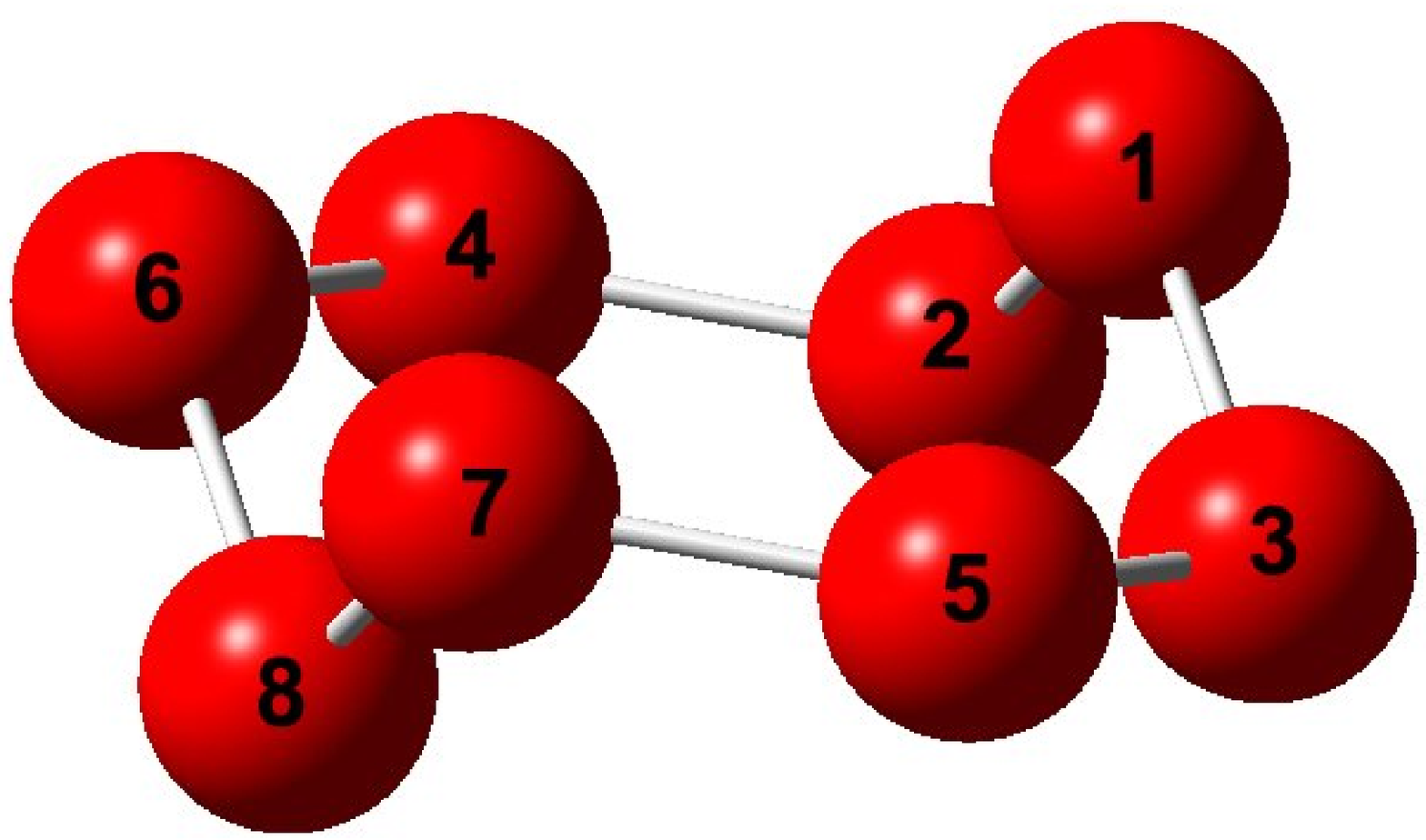}
\includegraphics[width=0.45\columnwidth]{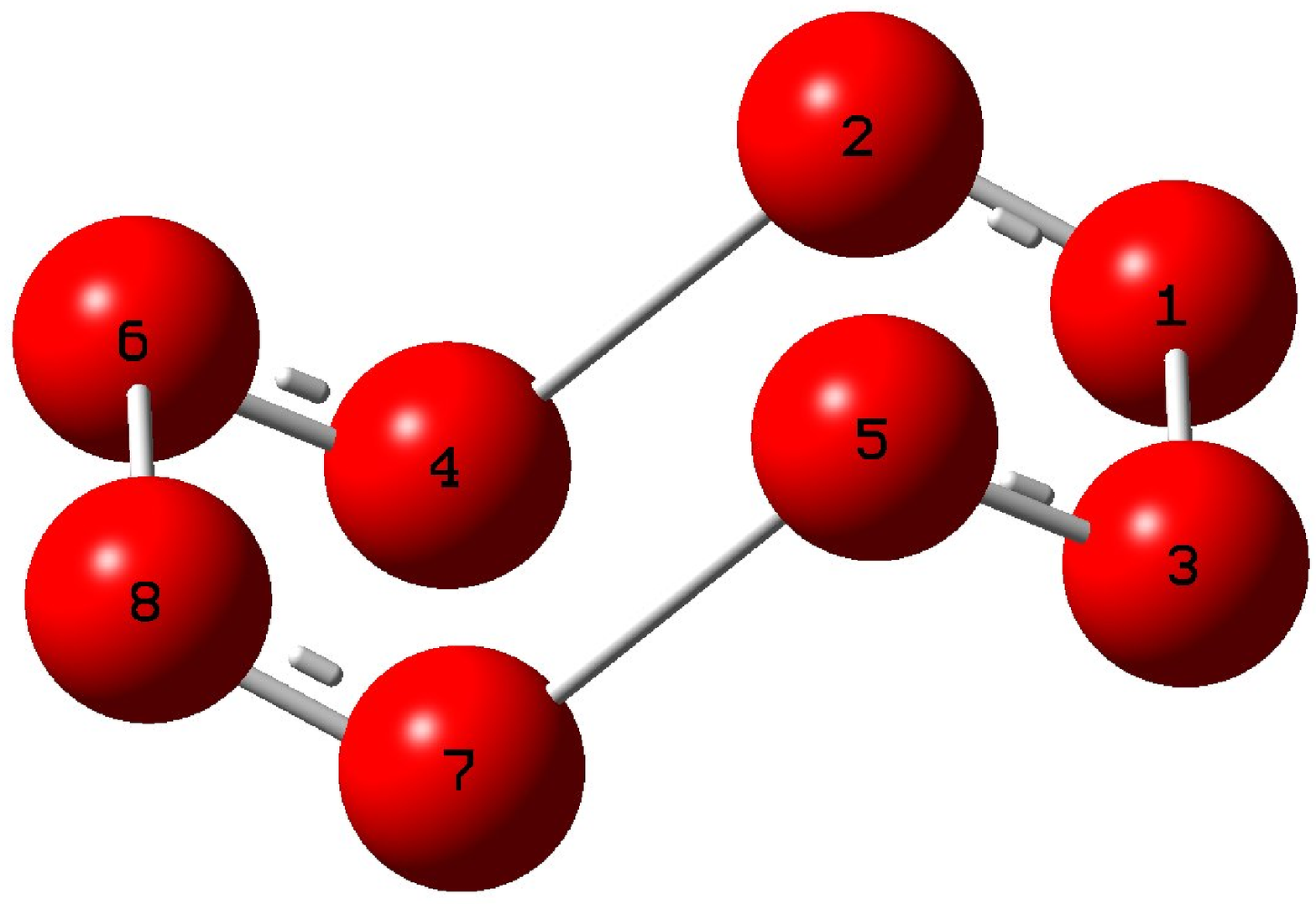}
\caption{Two low-lying isomers of O$_8$, both in the singlet
configuration: within HF+MP2 theory (left), and within PBE theory (right). See
Table~\ref{tab:structure} for structural details, and Table~\ref{tab:energies}
for ground-state and dissociation energy.}
\label{fig:2}
\end{figure}

\begin{figure}[t]
\centering
\includegraphics[height=0.8\columnwidth,angle=-90]{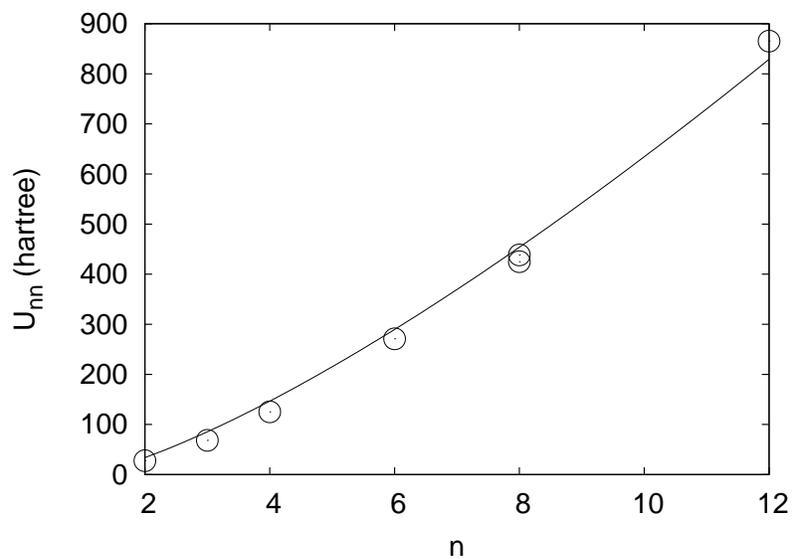}
\caption{Nuclear-nuclear repulsion energy, $U_{nn}$ (in hartrees), as a function
of the number of electrons $N$ in the low-lying isomers depicted in
Figs.~\ref{fig:1}, \ref{fig:2}, \ref{fig:3}, as well as for $n=2,3,4$. The two
points for $n=8$ refer to the two singlet clusters O$_8$ shown in
Fig.~\ref{fig:2}. The solid line is a guide to the eye, prompted by the proposal
of Mucci and March \cite{Mucci:79}.}
\label{fig:4}
\end{figure}

The authors acknowledge Professor C. Amovilli for providing relevant
data for Fig.~\ref{fig:2}.
NHM wishes to acknowledge that his contribution to the present article was made
during a visit to the University of Catania. NHM thanks Professors R. Pucci and
G. G. N. Angilella for their kind hospitality.

\bibliographystyle{mprsty}
\bibliography{a,b,c,d,e,f,g,h,i,j,k,l,m,n,o,p,q,r,s,t,u,v,w,x,y,z,zzproceedings,Angilella}

\clearpage
\clearpage

\begin{table}[t]
\centering
\begin{tabular}{rrrrrrr}
\multicolumn{3}{c}{bond lengths} & \multicolumn{4}{c}{bond angles} \\
\multicolumn{7}{l}{\hspace{-3em} O$_6$ (Fig.~\ref{fig:1}):} \\
\multicolumn{2}{c}{all bonds:} & 1.427 & \multicolumn{3}{c}{all angles:} & 103.85 \\
\multicolumn{7}{l}{\hspace{-3em} O$_8$ singlet (Fig.~\ref{fig:2}, left):} \\
1 & 2 & 1.422 & 1 & 2 & 4 &  100.46 \\
1 & 3 & 1.435 & 1 & 3 & 5 &  113.68 \\      
2 & 4 & 1.422 & 2 & 1 & 3 &  113.18 \\
3 & 5 & 1.435 & 2 & 4 & 6 &  113.18 \\
4 & 6 & 1.435 & 3 & 5 & 7 &  113.18 \\
5 & 7 & 1.422 & 4 & 6 & 8 &  113.68 \\
6 & 8 & 1.435 & 5 & 7 & 8 &  100.46 \\
7 & 8 & 1.422 & 6 & 8 & 7 &  113.18 \\
\multicolumn{7}{l}{\hspace{-3em} O$_8$ singlet (Fig.~\ref{fig:2}, right):} \\
1 & 2 & 1.203 & 1 & 2 & 4 & 109.878  \\ 
1 & 3 & 1.928 & 1 & 3 & 5 & 97.817   \\
2 & 4 & 1.854 & 2 & 1 & 3 & 97.794   \\
3 & 5 & 1.204 & 2 & 4 & 6 & 109.825  \\
4 & 6 & 1.203 & 3 & 5 & 7 & 109.776  \\
5 & 7 & 1.855 & 4 & 6 & 8 & 97.674   \\
6 & 8 & 1.934 & 5 & 7 & 8 & 109.887  \\
7 & 8 & 1.203 & 6 & 8 & 7 & 97.646   \\
\multicolumn{7}{l}{\hspace{-3em} O$_8$ singlet (CCSD/6-311++**):} \\
%{\bf
1 &  2 &  1.413 & 1 &  2 &  4 &   112.86 \\
1 &  3 &  1.413 & 1 &  3 &  5 &   112.86 \\	 
2 &  4 &  1.422 & 2 &  1 &  3 &   100.52 \\
3 &  5 &  1.422 & 2 &  4 &  6 &   113.36 \\
4 &  6 &  1.422 & 3 &  5 &  7 &   113.36 \\
5 &  7 &  1.422 & 4 &  6 &  8 &   112.86 \\
6 &  8 &  1.413 & 5 &  7 &  8 &   112.86 \\
7 &  8 &  1.413 & 6 &  8 &  7 &   100.52 \\
%}
\end{tabular}
\caption{Bond lengths (in \AA) and angles (in $^\circ$) for the clusters O$_6$
and O$_8$ considered in this work (see caption of Fig.~\ref{fig:3} for details
on the structure of O$_{12}$).
The last group of data refer to the structure of the unstable O$_8$
singlet cluster, obtained at the refined CCSD/6-311++** level of approximation.}
\label{tab:structure}
\end{table}

\begin{table}[t]
\centering
\begin{tabular}{r>{$}c<{$}>{$}r<{$}>{$}r<{$}l}
\hline
$n$ & S  & E(\mathrm{O}_n) & E(\mathrm{O}_n - n \mathrm{O})& \\
\hline
1  &      t  &	 -74.92178  &	0.00000   & \\
2  &      t  &	 -150.02987 &	-0.18631  & \\
3  &      s  &	 -224.99167 &	-0.22633  & \\
4  &      s  &	 -299.87402 &	-0.18690  & \\
6  &      s  &	 -449.90881 &	-0.37813  & Fig.~\ref{fig:1} \\
8  &      s  &     -599.85359 &     -0.47935  & Tab.~\ref{tab:structure} at CCSD \\
8  &      s  &     -600.70181 &     -1.32757  & Fig.~\ref{fig:2} at PBE \\
12 &      s  &	 -899.82570 &	-0.76434  & Fig.~\ref{fig:3} \\
%6 & $-449.989$ & $-0.377$ & Fig.~\ref{fig:1} \\
%8 & $-599.854$ & $-0.478$ & Fig.~\ref{fig:2} (left) \\
%12 & $-899.8257$ & $-0.761$ & Fig.~\ref{fig:3} \\
\end{tabular}
\caption{Ground-state and
dissociation energies (in a.u.) for the O$_n$ clusters ($n=6,8,12$) discussed in
the text. $S=s$ ($S=t$) refers to a singlet (triplet) spin
configuration, respectively. Relevant data for other atomic and molecular clusters of
interest (O$_n$, with $n=1,2,3,4$) are also included, for reference.}
\label{tab:energies}
\end{table}

\end{document}